\documentstyle[epsfig]{aipproc}

\newcommand{\ol}{\overline}
\newcommand{\Pslash}{\kern 0.2 em P\kern -0.56em \raisebox{0.3ex}{/}}
\newcommand{\pslash}{\kern 0.2 em p\kern -0.4em /}
\newcommand{\kslash}{\kern 0.2 em k\kern -0.45em /}
\newcommand{\Sslash}{\kern 0.2 em S\kern -0.56em \raisebox{0.3ex}{/}}
\newcommand{\Mslash}{\kern 0.2 em M\kern -0.70em \raisebox{0.3ex}{/}}
\newcommand{\g}{\gamma}
\newcommand{\sig}{\sigma}
\newcommand{\eps}{\epsilon}
\newcommand{\dg}{\dagger}

\newcommand{\sT}{{\scriptscriptstyle T}}
\newcommand{\nn}{\nonumber}
\newcommand{\newangle}{{<\kern -0.3 em{\scriptscriptstyle )}}}

\begin{document}
\title{Calculation of Fragmentation Functions \\ in Two-hadron
Semi-inclusive Processes}

\author{A. Bianconi$^a$, S. Boffi$^b$, D. Boer$^c$, R. Jakob$^d$ \\ and 
\underline{M. Radici}$^b$}

\address{$^a$ Dipartimento di Chimica e Fisica per i Materiali e per
l'Ingegneria, Universit\`a di Brescia, \\ I-25133 Brescia, Italy\\
$^b$ Dipartimento di Fisica Nucleare e Teorica, Universit\`{a} di 
Pavia, and\\
Istituto Nazionale di Fisica Nucleare, Sezione di Pavia, 
I-27100 Pavia, Italy\\
$^c$ RIKEN/BNL Research Center, Physics Department, NY 11973 Upton, USA \\
$^d$ University of Wuppertal, Theoretical Physical Department, D-42097
Wuppertal, Germany}

\maketitle

\begin{abstract}
We investigate the properties of interference fragmentation functions 
arising from the emission of two leading hadrons inside the same jet for 
inclusive lepton-nucleon deep-inelastic scattering. Using an extended 
spectator model for the mechanism of the hadronization, we give a complete
calculation and numerical estimates for the examples of a proton-pion pair 
produced with invariant mass on the Roper resonance, and of two pions
produced with invariant mass close to the $\rho$ mass. We discuss
azimuthal angular dependence of the leading order cross section to point
up favourable conditions for extracting transversity from experimental
data.
\end{abstract}

\section*{Introduction}
\label{S:mradici:intr}

Because of the still lacking rigorous explanation of confinement, the 
nonperturbative nature of quarks and gluons inside hadrons can be explored
by extracting information from distribution (DF) and fragmentation 
functions (FF) in hard scattering processes. There are three fundamental 
DF that completely characterize the quark inside hadrons at leading 
twist with respect to its longitudinal momentum and spin: the momentum 
distribution $f_1$, the helicity distribution $g_1$ and the transversity 
distribution $h_1$. At variance with the first two ones, $h_1$ is 
difficult to address because of its chiral-odd nature. A complementary 
information can come from the analysis of the hadrons produced by the 
fragmentation process of the final quark, namely from FF. So far, only the 
leading unpolarized FF, $D_1$, is partly known, which is the counterpart 
of $f_1$. The basic reason for such a poor knowledge is related to the 
difficulty of measuring more exclusive channels in hard processes. The 
new generation of machines (HERMES, COMPASS, RHIC) and planned projects
(ELFE, EPIC) allow for a much more powerful final-state identification 
and, therefore, for a wider and deeper analysis of FF, particularly when 
Final State Interactions (FSI) are considered. In this context, naive 
``T-odd'' FF naturally arise because the existence of FSI prevents 
constraints from time-reversal invariance to be applied to the 
fragmentation process~\cite{mradici:vari1}. This new set of FF includes 
also chiral-odd objects that become the natural partner needed to isolate 
$h_1$.

The presence of FSI allows that in the fragmentation process there are at
least two competing channels interfering through a nonvanishing phase. 
However, this is not enough to generate naive ``T-odd'' FF. Excluding
{\it ab initio} any mechanism breaking factorization, there are 
basically two ways to describe the residual interactions of the leading 
hadron inside the jet: assume the hadron moving in an external effective 
potential, or model microscopically independent interaction vertices that 
lead to interfering competing channels. In the former case, introduction 
of an external potential in principle breaks the translational and 
rotational invariance of the problem. Further assumptions can be made 
about the symmetries of the potential, but at the price of loosing 
interesting contributions to the amplitude such as those coming from naive 
``T-odd'' FF~\cite{mradici:prd1}. In the latter case, the difficulty 
consists in modelling a genuine interaction vertex that cannot be 
effectively reabsorbed in the soft part describing the hadronization. This 
poses a serious difficulty in modelling the quark fragmentation into one 
observed hadron because it requires the ability of modelling the FSI 
between the hadron itself and the rest of the jet~\cite{mradici:prd1}. 
Therefore, here we will consider a hard process, semi-inclusive 
Deep-Inelastic Scattering (DIS), where the hadronization leads to two 
observed hadrons inside the same jet. A new set of interference FF arise at 
leading twist~\cite{mradici:prd1} and their symmetry properties are
briefly reviewed. For the hadron pair being a proton and a pion with 
invariant mass equal to the Roper resonance, we have already estimated
these FF using an extended version of the diquark spectator 
model~\cite{mradici:prd2}. In this case, FSI come from the interference 
between the direct production of the two hadrons and the decay of the 
Roper resonance. Here, we present results for the hadron pair being two 
pions with invariant mass around the $\rho$ mass.

\section*{Quark-quark correlation function}
\label{S:mradici:FF}

\begin{figure}[h]
\begin{center}
\psfig{file=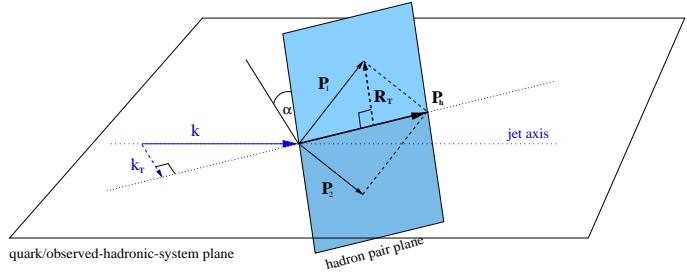, width=9cm}
\end{center}
\caption{Kinematics for a fragmenting quark jet containing a pair of 
leading hadrons.}
\label{F:mradici:1}
\end{figure}

In analogy with semi-inclusive hard processes involving one detected 
hadron in the final state~\cite{mradici:piet96}, the simplest matrix 
element for the hadronisation into two hadrons is the quark-quark 
correlation function describing the decay of a quark with momentum $k$ 
into two hadrons $P_1, P_2$, namely  
\begin{equation}
\Delta_{ij}(k;P_1,P_2)= \displaystyle{\sum_X} \int
\frac{d^4\zeta}{(2\pi)^4} \; 
e^{ik\cdot\zeta}\langle 0|\psi_i(\zeta)\,a_{P_2}^\dg\,a_{P_1}^\dg |X
\rangle \; \langle X| a_{P_1}\,a_{P_2}\,\ol{\psi}_j(0)|0\rangle \;,
\label{E:mradici:Delta}
\end{equation}
where the sum runs over all the possible intermediate states involving the 
two final hadrons $P_1,P_2$. Since the three external momenta $k,P_1,P_2$ 
cannot all be collinear at the same time, we choose for convenience the 
frame where the total pair momentum $P_h=P_1+P_2$ has no transverse 
component. By generalizing the Collins-Soper light-cone 
formalism~\cite{mradici:colsop} for fragmentation into multiple hadrons, 
the cross section for two-hadron semi-inclusive emission is a linear
combination of projections $\Delta^{\left[ \Gamma \right]}$ of $\Delta$ by
specific Dirac structures $\Gamma$, after integrating over the 
(hard-scale suppressed) light-cone component $k^+$ and, consequently, 
taking $\zeta$ as light-like~\cite{mradici:prd1}. At leading order, we get
\begin{eqnarray} 
\Delta^{[\g^-]}(z_h,\xi,{\bf k}_\sT^2,{\bf R}_\sT^2,
{\bf k}_\sT \cdot {\bf R}_\sT) &\equiv& D_1 \nn \\
\Delta^{[\g^- \g_5]}(z_h,\xi,{\bf k}_\sT^2,{\bf R}_\sT^2,{\bf k}_\sT \cdot 
   {\bf R}_\sT) &\equiv& 
     \frac{\eps_\sT^{ij} \,R_{Ti}\,k_{Tj}}{M_1\,M_2}\; G_1^\perp \nn \\
\Delta^{[i\sig^{i-} \g_5]}(z_h,\xi,{\bf k}_\sT^2,{\bf R}_\sT^2,
{\bf k}_\sT \cdot {\bf R}_\sT) &\equiv& {\epsilon_\sT^{ij}R_{Tj}\over 
M_1+M_2}\, H_1^{\newangle}+{\epsilon_\sT^{ij}k_{Tj}\over M_1+M_2}\,
 H_1^\perp \; , 
 \label{E:mradici:FF} 
\end{eqnarray}  
where $\epsilon_\sT^{\mu\nu} = \epsilon^{-+\mu\nu}$. The functions 
$D_1,G_1^\perp,H_1^{\newangle},H_1^\perp$ are the interference FF that
depend on how much of the fragmenting quark momentum $k$ is carried by the 
hadron pair $(z_h=z_1+z_2)$, on the way this momentum is shared inside the 
pair $(\xi=z_1/z_h)$, and on the ``geometry'' of the pair, namely on the
transverse relative momentum of the two hadrons $({\bf R}_\sT^2)$ and on 
the relative orientation between the pair plane and the quark jet axis 
$({\bf k}_\sT^2,{\bf k}_\sT \cdot {\bf R}_\sT$, see also
Fig.~\ref{F:mradici:1}).
The different Dirac structures $\Gamma$ are related to 
different spin states of the fragmenting quark and lead to the nice 
probabilistic interpretation at leading order~\cite{mradici:prd1}: 
$D_1$ is the probability for an unpolarized quark to produce a pair of 
unpolarized hadrons; $G_1^\perp$ is the difference of probabilities for a 
longitudinally polarized quark with opposite chiralities to produce a pair 
of unpolarized hadrons; $H_1^{\newangle}$ and $H_1^\perp$ both are 
differences of probabilities for a transversely polarized quark with 
opposite spins to produce a pair of unpolarized hadrons. $G_1^\perp$, 
$H_1^{\newangle}$ and $H_1^\perp$ are (naive) ``T-odd'' and do not vanish 
only if there are residual interactions in the final state. In this case, 
the constraints from time-reversal invariance cannot be applied. 
$G_1^\perp$ is chiral even; $H_1^{\newangle}$ and $H_1^\perp$ are chiral 
odd and can, therefore, be identified as the chiral partners needed to 
access the transversity $h_1$. Given their probabilistic interpretation, 
they can be considered as a sort of ``double'' Collins 
effect~\cite{mradici:coll}.

\section*{Numerical results}
\label{S:mradici:out}

\begin{figure}[hbtp]
\begin{center}
\psfig{file=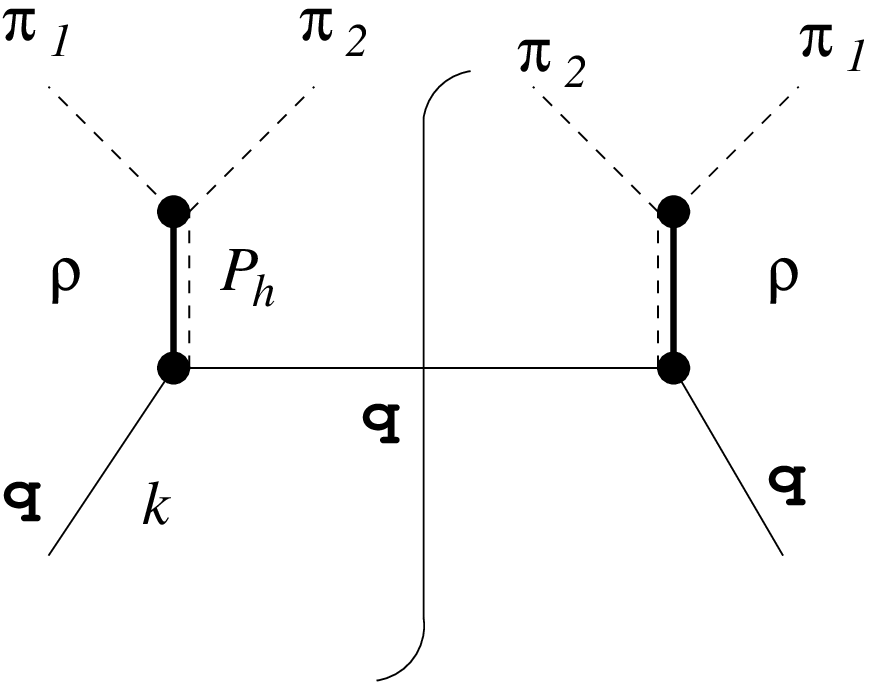, width=3.5cm}
\hspace{5mm}
\psfig{file=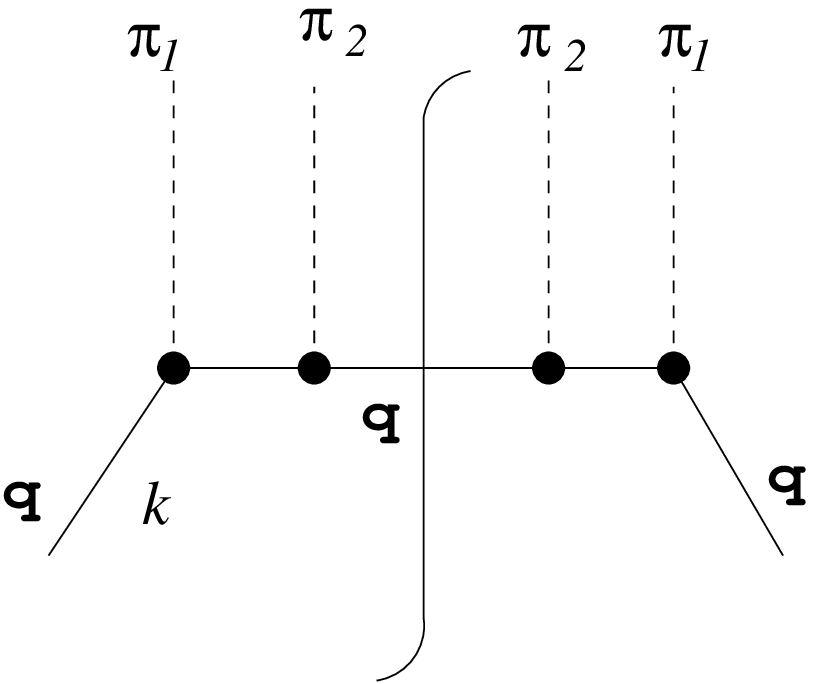, width=2.8cm}
\hspace{5mm}
\psfig{file=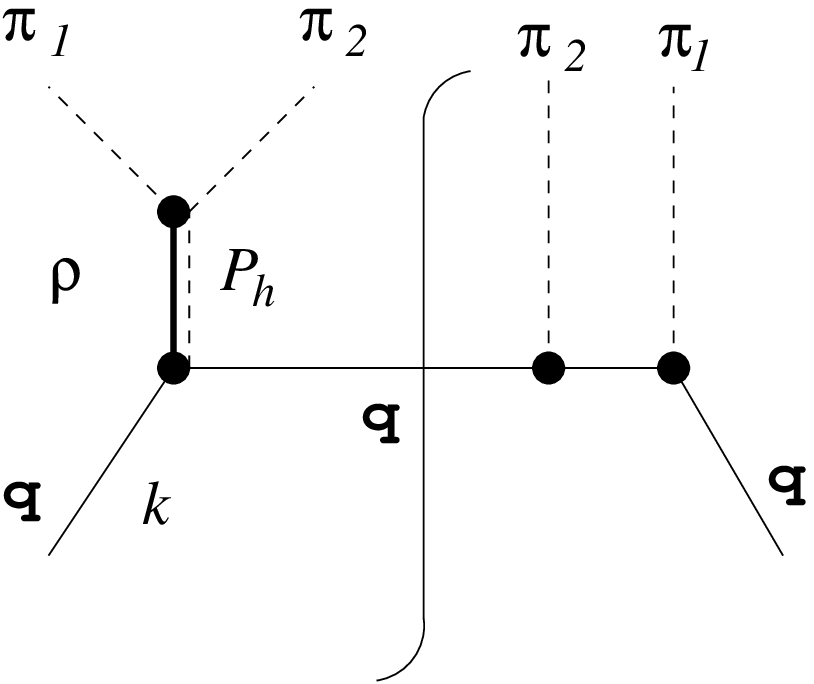, width=3cm}\\
\vspace{2mm}
$a$ \hspace{35mm} $b$ \hspace{35mm} $c$
\end{center}
\caption{\label{F:mradici:3}Diagrams for quark $q$ decay into two pions 
through a direct channel or a $\rho$ resonance.}
\end{figure}

\begin{figure}[h] 
\begin{center}
\epsfig{file=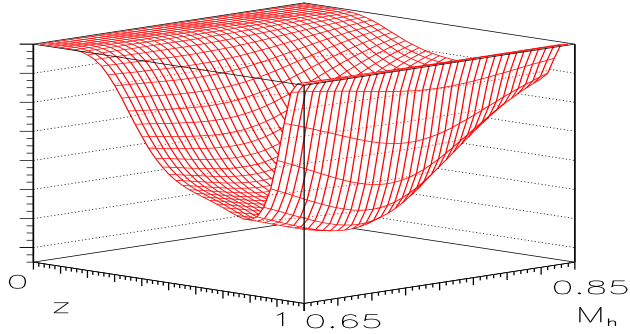, width=9cm,
height=5cm}
\end{center}
\caption{$H_1^{\newangle}(z_h,M_h)$ for the fragmentation of a quark $u$ 
into $\pi^+$ and $\pi^-$.}
\label{F:mradici:4}
\end{figure}

In order to make quantitative predictions, we adopt the formalism of the 
spectator model, specializing it to the emission of a hadron pair. The 
basic idea is to replace the sum over the complete set of intermediate 
states in Eq.~(\ref{E:mradici:Delta}) with an effective spectator state 
with a definite mass $M_D$, momentum $P_D$. Consequently, the correlator 
simplifies to 
\begin{equation} 
\Delta_{ij}(k;P_1,P_2)\sim
\frac{\theta(P_D^+)}{(2\pi)^3} \; \delta\left((k-P_h)^2-M_D^2\right)\;
\langle 0|\psi_i(0)|P_1,P_2,D\rangle\langle D,P_2,P_1|\ol{\psi}_j(0)
|0\rangle \;,
\label{E:mradici:specDelta}
\end{equation}
where the additional $\delta$ function allows for a completely analytical
calculation of the Dirac projections (\ref{E:mradici:FF}). For the hadron
pair being a proton and a pion with invariant mass the mass of the Roper
resonance, results have been published in Ref.~\cite{mradici:prd2}. In
this case, the spectator state has the quantum numbers of a scalar or
axial diquark. FSI arise from the interference between the direct
production and the decay of the Roper resonance. Here, we show results for
the hadron pair being two pions with invariant mass in the range $\left[
m_{\rho}-\Gamma_{\rho}, m_{\rho}+\Gamma_{\rho} \right]$, $m_{\rho}=768$
MeV and $\Gamma_{\rho} \sim 250$ MeV. The spectator 
states becomes an on-shell quark with mass $m_q=340$ MeV. The quark 
decay is specialized to the set of diagrams shown in 
Fig.~\ref{F:mradici:3}, and their hermitean conjugates, where the naive
``T-odd'' FF now arise from the interference between the direct production
of the two $\pi$ and the decay of the $\rho$. 
Diagram~\ref{F:mradici:3}a accounts for almost all the strength of
$\pi-\pi$ production in the relative $P$-channel. We have explicitly
checked that diagram~\ref{F:mradici:3}b reproduces the experimental
transition probability for $\pi-\pi$ production in the relative 
$S$-channel. Hence, we believe this choice represents most of the
$\pi-\pi$ strength for invariant mass in the considered interval. We
define Feynman rules for the $\rho \pi \pi$, $q\pi q$ and $q\rho q$
vertices introducing cut-offs to exclude large virtualities of the quark 
while keeping the asymptotic behaviour of FF at large $z_h$ consistent 
with the quark counting rule. We infer the vertex form factors from 
previous works on the spectator model~\cite{mradici:jak97}.
However, there numbers should be taken as indicative, since the ultimate
goal is to verify that nonvanishing ``T-odd'' FF occur, particularly when
integrating on some of the kinematical variables and possibly washing all
interference effects out. Results of analytical calculation of 
Eq.(\ref{E:mradici:specDelta}) show that $H_1^{\perp}=0$ and 
$H_1^{\newangle}=-2m_q G_1^{\perp}/m_{\pi}$. After integrating over 
${\bf k}_\sT, {\bf R}^2_\sT$ while keeping ${\bf R}_\sT$ in the horizontal
plane of the lab (usually identified with the scattering plane), we still
get nonvanishing FF. As an example, $H_1^{\newangle}(z_h,M_h)$ is shown in
Fig.~\ref{F:mradici:4} for the fragmentation $u\rightarrow \pi^+ \pi^-$, 
where $M_h$ is the invariant mass of the two pions.
\vspace{-.5cm}
The cross section for the deep-inelastic scattering of an unpolarized
electron on a polarized proton target where two pions are detected in the 
final state, contains, after integrating all the transverse dynamics 
(${\bf P}_{h T}, {\bf k}_\sT, {\bf R}^2_\sT$), an unpolarized 
contribution proportional to $D_1(z_h,M_h)$ and a term proportional to
$H_1^{\newangle}(z_h,M_h)$ which depends on the transverse target
polarization $S_T$. Therefore, by flipping the polarization of the target,
it is possible to build the following azimuthal asymmetry
\begin{equation}
{\cal A} \propto \frac{S_T}{2m_{\pi}} \, \sin (\phi_{R_T}+\phi_{S_T}) \,
\frac{h_1(x)}{f_1(x)} \, \frac{H_1^{\newangle}(z_h,M_h)}{D_1(z_h,M_h)} 
\; ,
\label{E:mradici:asum}
\end{equation}
where $\phi_{R_T}, \phi_{S_T}$ are the azimuthal angles of ${\bf R}_\sT,
{\bf S}_\sT$ with respect to the scattering plane, respectively. The
asymmetry shows indeed the familiar sinusoidal azymuthal dependence. 
Noteworthy is the factorization of the 
chiral-odd, naive ``T-odd'' $H_1^{\newangle}$ from the chiral-odd 
transversity $h_1$. Therefore, such asymmetry measurement allows for the
extraction of $h_1$ using a model input for the FF.

\end{document}